\documentclass[%
 reprint,
 amsmath,amssymb,
 aps,
 pra,
]{revtex4-2}

\usepackage{natbib}
\usepackage{amsfonts}
\usepackage{graphicx}
\usepackage{graphicx, nicefrac}
\usepackage{dcolumn}
\usepackage{float}
\usepackage{bm}
\def\abs#1{\left| #1 \right|}

\usepackage{braket} 
\usepackage{float}
\begin{document}

\preprint{APS/123-QED}

\title{
Three-laser coherent population trapping in a multi-$\Lambda$ system: Theory, experiment, and applications}

\author{Nicol\'{a}s A. Nu\~{n}ez Barreto}
\author{Mart\'{i}n Drechsler}%
\author{Christian T. Schmiegelow}
 \email{schmiegelow@df.uba.ar}
\affiliation{%
 Laboratorio de Iones y \'{A}tomos Fr\'{i}os, Departamento de F\'{i}sica, FCEyN, UBA, Pabell\'{o}n 1, Ciudad Universitaria, 1428 Buenos Aires, Argentina and Instituto de F\'{i}sica de Buenos Aires, UBA CONICET, Pabell\'{o}n 1, Ciudad Universitaria, 1428 Buenos Aires, Argentina
}%

\date{\today}

\begin{abstract}
We present theoretical and experimental results of coherent population trapping spectra on a multilevel $\Lambda$-type configuration, adding a third beam to the standard two-laser $\Lambda$ system to avoid undesired optical pumping. We show that the extra laser can preserve the nature of the dark resonances or introduce decoherence depending on its power. Experiments are carried out using a single trapped $^{40}$Ca$^+$ ion in the $S_{1/2}-P_{1/2}-D_{3/2}$ manifold. Theoretically, the problem is solved with a Floquet-type expansion of the Liouvillian that correctly predicts all of the measured spectra without the need of full time integration. As a first application of the multilaser technique, we show that the richer spectra obtained can be used as a vectorial polarimeter of one of the beams, allowing one to measure the electrical field at the ion’s position in any spatial direction. We also explain how our setup could realize a thermometer with tunable sensitivity and no laser-linewidth dependence.
\end{abstract}

\maketitle

Coherent population trapping (CPT)~\cite{arimondo1976nonabsorbing,gray1978coherent,arimondo1996v} and electromagnetically induced transparency (EIT)~\cite{harris1990nonlinear,fleischhauer2005electromagnetically} occurring in three-level $\Lambda$-type atomic systems reveal dark resonances with spectral linewidths much finer than the natural linewidths of the excited transitions. In fact, these are only fundamentally limited by dephasing and decay into the lower intervening states, which are normally considered stable or at least metastable, with lifetimes typically beyond 1~s if not days or years. 

These two similar phenomena, which differ only in the relative strengths of the involved laser fields, have been widely observed and studied in diverse systems, for example, in cold-atomic gases~\cite{boller1991observation} and Bose-Einstein condensates~\cite{ahufinger2002electromagnetically}. CPT spectroscopy is also a promising perspective in the area of quantum metrology and quantum clocks with neutral atom vapor cells~\cite{vanier2005atomic,lenci2009dark,shah2010advances,warren2017experimental}. 
Since EIT generates strong nonlinear effects, these kinds of resonances find applications in the field of quantum information as a mechanism to slow down light~\cite{hau1999light,kash1999ultraslow} and, therefore, to store optical pulses~\cite{liu2001observation,phillips2001storage,ginsberg2007coherent}. It can also enhance nonlinear optical effects such as sum-frequency generation ~\cite{zhang1993nonlinear,zhang1995sum,jain1996efficient} and four-wave mixing~\cite{hemmer1995efficient}.

More recently, several experiments of CPT and EIT were performed with trapped ions. The narrow feature of the resonances is especially useful to study the motional states of few ions because the resonances widths are especially sensitive to ions' motion~\cite{schubert1989line,reiss2002raman,lisowski2005dark}. This makes it possible to use it to measure the temperature of single trapped ions~\cite{rossnagel2015fast,tugaye2019absolute} as well as using EIT to perform cooling below the Doppler limit of single ions~\cite{morigi2000ground,roos2000experimental} and even multimode cooling of long ion chains and crystals~\cite{lin2013sympathetic,lechner2016electromagnetically,scharnhorst2018experimental}. This technique proved to be a robust cooling method towards coherent control of multiple ions simultaneously with applications in quantum computing and quantum simulation~\cite{schneider2012experimental,bruzewicz2019trapped}. 

\begin{figure*}[ht]
\includegraphics[scale=1]{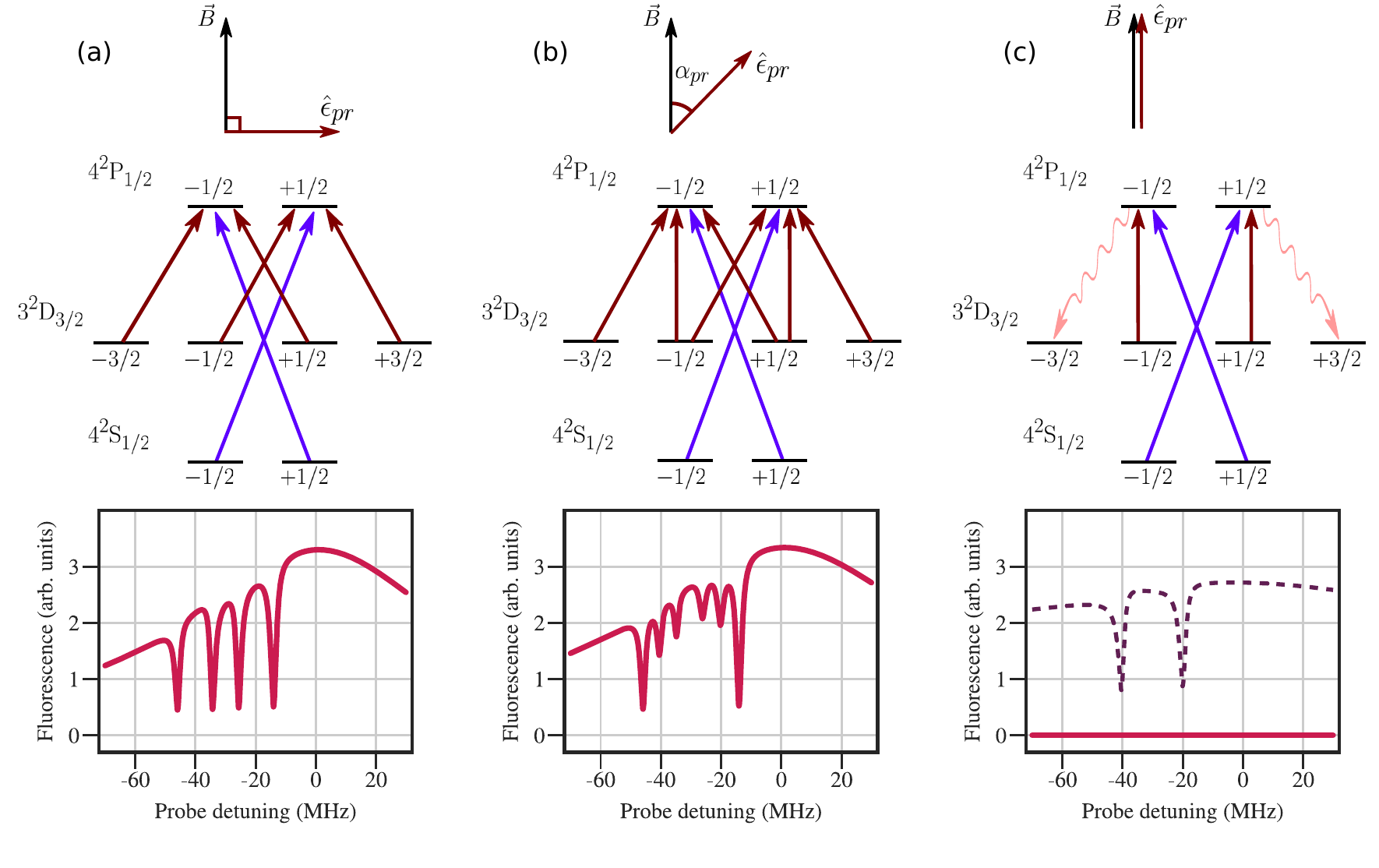}
\caption{\label{fig:01_cpttheo} Energy levels of $^{40}$Ca$^+$ ion with the allowed transitions and its corresponding dark resonance spectra for three different configurations of linear probe polarizations angle $\alpha_{\textrm{pr}}$ with respect to $\vec{B}$. For all plots, $\hat{\epsilon}_{\textrm{dop}}=\hat{\sigma}_+ + \hat{\sigma}_-$ was set. (a) For $\hat{\epsilon}_{\textrm{pr}}=\hat{\sigma}_+ + \hat{\sigma}_-$, i.e., $\alpha_{\textrm{pr}} = 90^{\circ}$, the spectrum shows four dark resonances. (b) When $0^{\circ}<\alpha_{\textrm{pr}}<90^{\circ}$, the probe spectrum shows six dark resonances, and the strengths of them strongly depend on $\alpha_{\textrm{pr}}$. (c) For $\hat{\epsilon}_{\textrm{pr}}=\hat{\pi}$, which means $\alpha_{\textrm{pr}}=0^{\circ}$, the spectrum should show two dark resonances. But since this polarization generates optical pumping to the $m_j=\pm 3/2$ $D$ sublevels, the stationary fluorescence is zero and there is not an observable spectrum (solid line), thereby hindering the two dark resonances. If we turn on an additional laser in the $D$-$P$ transition, a repumper, with polarization $\hat{\epsilon}_{\textrm{rep}}=\hat{\sigma}_+ + \hat{\sigma}_-$ and blue-detuned, the two $\hat{\pi}$ dark resonances of the spectrum can be seen (dashed line).}
\end{figure*}

Pure three-level systems rarely occur in nature. However, typical atomic systems with multi-level configurations can also exhibit dark resonances. In such cases, to produce a closed three- or more-level $\Lambda$-type system, a proper combination of laser polarization and geometry must be chosen. Otherwise, open decay pathways can lead to optical pumping into states that do not fluoresce, which hinders the observation of the dark resonances. 


In this paper we study in detail the role of an extra repumper as a method to avoid optical pumping into dark states when observing CPT with a single trapped ion. Similar work has been carried out with atomic vapours~\cite{renzoni1997coherent,dey2015study,li2022repumping}. Here we show how introducing this additional laser considering its coherent effects on the system will either preserve the resonances or dephase them, depending on its power. This can be especially useful as it allows the observation of the full dark resonance spectrum irrespective of the polarization of the probe beam.

Also, we present two prospective applications of these three-beam CPT resonances. The main one concerns three-dimensional (3D) polarimetry of vector beams. For this application we take advantage of the fact that the shape, position, and strength of the dark resonances that appear in a spectrum strongly depend on the relation between the polarization of the beam and the magnetic field direction. Thus, the observed spectra provides information of the 3D polarization of the electric fields of the intervening beams, similarly to the recent demonstration of the measurement of the 3D magnetic field alignment using an atomic cloud and a vector vortex beam~\cite{castellucci2021atomic}. As the typical size of the wave packet of a trapped ion is in the 10-100~nm range, it can be used as a 3D polarimeter with spatial resolution below the diffraction limit of strongly focused beams, allowing to map the polarization distribution of focused beams. 

The paper is organized as follows. In Sec. I we introduce the different scenarios where the dark resonance spectra appear in a multi-level $\Lambda$-type system and discuss the addition of a third laser to avoid optical pumping. In Sec. II we develop the formalism needed to calculate CPT spectra for two and three lasers and compare the solutions. In Sec. III we show experimental results of three-laser CPT spectra for different cases and analyze the effects of the third laser for the low and high saturation limits. In Sec. IV we present two methods for doing polarimetry of one of the beams using the dark resonances. Finally, in Sec. V we discuss the use of the dark resonances to perform thermometry of trapped ions.

\section{INTRODUCTION}

Through all this work we will focus on $^{40}$Ca$^+$ level system, but our analysis can be extended to other atomic systems. In particular we study the multi-$\Lambda$ system formed by the ground state $4^2$S$_{1/2}$, one excited state $4^2$P$_{1/2}$, and the metastable state $3^2$D$_{3/2}$, as seen in Fig.~\ref{fig:01_cpttheo}. The $S$-$P$ transition near $397$~nm and the $D$-$P$ transition near $866$~nm are dipole connected with the excited short-lived level which has a lifetime of $6.9$~ns \cite{hettrich2015measurement}. When exciting the atom with two nearly resonant lasers, one for each transition, but an equal detuning with respect to the $P$ level, a dark resonance occurs. However, these dark resonances may be hindered by optical pumping under certain conditions, as we describe below.

As a quick rule of thumb, which would be useful to keep in mind for the present discussion, one can consider the following cases for the selection rules. Considering the quantization direction as the one given by the magnetic field, the three possible transitions are:  $\pi$ ($\Delta m_j = 0)$, which occurs when the polarization is parallel to $\vec{B}$, $\sigma_+$ ($\Delta m_j = +1$) or $\sigma_-$ ($\Delta m_j = -1$) which takes place when the polarization is right or left circularly polarized in the plane perpendicular to $\vec{B}$. Also, if the laser is linearly polarized  orthogonally to $\vec{B}$, both $\sigma_{+}$ and $\sigma_-$ can occur, as the polarization can be decomposed as a superposition of both circular polarizations. Throughout this work we will name the polarizations of the lasers with these labels, according to the possible transitions they can drive.

With only two lasers, some configurations give the expected fluorescence spectra with dark resonances, while some others lead to optical pumping which hinders fluorescence and consequently the observation of dark resonances, as seen in Fig.~\ref{fig:01_cpttheo}. There, the $397$-nm laser (\textit{Doppler} laser) is kept at fixed detuning to the red of the transition with linear polarization perpendicular to the magnetic field, such that $\hat{\epsilon}_{\textrm{dop}}=\hat{\sigma}_+ + \hat{\sigma}_-$. The spectrum is obtained by sweeping the frequency of a second beam near $866$~nm (\textit{probe} laser). The polarization of this beam is always kept linear but its direction can be changed between different spectrum measurements.

We show in Fig. 1 three examples of dark resonance spectra of the probe laser, for three different polarizations $\hat{\epsilon}_{\textrm{pr}}$. These are retrieved by fixing the detuning of the Doppler laser and computing the fluorescence, proportional to the population of the excited states, while varying the detuning of the probe. In the spectra, the number and location of the dark resonances will change depending on which three-level systems are excited by the two lasers, which will vary depending on the selection rules. In Fig. \ref{fig:01_cpttheo}(a), we set the polarization of the probe laser to be $\hat{\epsilon}_{\textrm{pr}}=\hat{\sigma}_+ + \hat{\sigma}_-$, i.e., orthogonal to $\vec{B}$. With this choice, the spectrum shows four dark resonances which are separated because the Land\'{e} factors of the $S$ and $D$ levels are different. Two of them belong to $\sigma_+$ $D$-$P$ transitions and the other two correspond to $\sigma_-$ $D$-$P$ transitions.
If we reduce the angle between the probe polarization and the magnetic field $\alpha_{\textrm{pr}}$ to some value between $0^\circ$ and $90^\circ$, as can be seen in Fig. \ref{fig:01_cpttheo}(b) for $\alpha_{\textrm{pr}}=15^\circ$, two additional dark resonances emerge. These two dark resonances correspond to $\pi$ transitions and add up to the previous four, conforming a total of six dark resonances that can appear in this lasers geometry. It can also be seen that the depths of the dark resonances change significantly with $\alpha_{\textrm{pr}}$. This shows that the dark resonances are especially sensitive to the probe polarization. 
Finally, in Fig. \ref{fig:01_cpttheo}(c) we show what happens when the polarization of the probe is parallel to $\vec{B}$, i.e., $\hat{\epsilon}_{\textrm{pr}}=\hat{\pi}$. 
The emitted fluorescence in the stationary state is zero for all probe detunings since optical pumping to $m_j = \pm 3/2$ $D$ magnetic sublevels occur. Therefore, the two dark resonances that should appear, plotted in dashed line, do not show up. The same behavior is expected if the probe polarization is chosen purely $\hat{\sigma}_+$ or $\hat{\sigma}_-$. This implies that it is not possible to observe dark resonances for all possible probe polarizations by doing CPT with two lasers. To do this, one should find a method to avoid optical pumping to dark states. One possibility is to add a second laser in the $D$-$P$ transition, a \textit{repumper}, with fixed detuning and polarization such that all the $D$ sublevels are repumped, for example $\hat{\sigma}_+ + \hat{\sigma}_-$. This laser will repump the population from all $D$ magnetic sublevels thus avoiding population trapping and therefore allowing one to observe the spectrum for all possible polarizations of the probe laser.

In the next Sec. we will present the theory needed to consider a third laser, the \textit{repumper}, in the atomic dynamics and the methods to retrieve the atomic spectra. After that, we will show experimental results of the model and describe two applications of this, as a vectorial polarimeter and as an ion thermometer.

\section{CPT THEORY}
We will now develop the formalism to calculate the atomic spectra, focused particularly in the probe spectrum. We begin by making an overview of the well-known standard CPT theory with two lasers. After that, we show the modifications needed to add a third laser to the system, the repumper laser. We consider this new laser to be tuned near the $866$-nm transition with a detuning and polarization independent of the probe laser. However, the same formalism can be extended to add a laser in the other transition, as well as more than one extra laser.  As the problem is inherently time-dependent, we show a formalism to calculate a time-averaged solution, which allows us to calculate the steady states of the system for any configuration. 

\subsection{Standard CPT theory}
The Hamiltonian of the atomic system interacting with the electric fields is
\begin{equation}
    H = H_{\textrm{atom}} + H_{\textrm{int}}. 
\end{equation}
There, $H_{\textrm{atom}}$ is the atomic Hamiltonian, and $H_{\textrm{int}}$ is the semiclassical Hamiltonian of the resonant dipolar interaction of the atom with the electric field of each laser. This last one can be written as
\begin{equation}
    H_{\textrm{int}} = -\vec{D} \cdot \sum_m \vec{E}_m e^{-i \omega_m t},
    \label{intham}
\end{equation}
where $\vec{D}$ is the atomic dipole operator which multiplies the sum of all the electric fields $\vec{E}_m$ of the lasers involved, with frequencies $\omega_m$. From this expression it is clear that the interaction Hamiltonian introduces an explicit time-dependency to the full Hamiltonian. To overcome this, the most common method used is to transform into a rotating frame with an unitary transformation $U$~\cite{oberst1999resonanceth}.

To write $U$ explicitly it is useful to name the energy levels with numbers from 1 to 8 considering the two $S$ ($\ket{1}$, $\ket{2}$), two $P$ ($\ket{3}$, $\ket{4}$) and four $D$ ($\ket{5}$, $\ket{6}$, $\ket{7}$, $\ket{8}$) magnetic Zeeman sublevels. In the case of two lasers, the appropriate transformation to use is
\begin{equation}
\begin{split}
U = e^{-i\omega_{\textrm{dop}} t}\left(  \ket{1}\bra{1}+\ket{2}\bra{2}\right)+\\ \ket{3}\bra{3}+\ket{4}\bra{4}+ \\
e^{-i\omega_{\textrm{pr}} t} (\ket{5}\bra{5}+...+\ket{8}\bra{8}),
\end{split}
\end{equation}
where $\omega_{\textrm{dop}}$ and $\omega_{\textrm{pr}}$ are the frequencies of the Doppler and the probe lasers respectively. This rotation leads to a time-independent Hamiltonian $\mathcal{H}_0=UHU^{\dagger}-i\hbar U\frac{dU^{\dagger}}{dt}$. Then, the dynamics of the system characterized by a density matrix $\rho$ is governed by a master equation given by
\begin{equation}
    \frac{d\rho}{dt} = -\frac{i}{\hbar}[\mathcal{H}_0, \rho] + \mathcal{L}_{\textrm{damp}}(\rho).
    \label{mastereq}
\end{equation}
There, $\mathcal{L}_{\textrm{damp}}$ is a Lindblad operator which can account for decoherence or decays in the system. We consider two kinds of decoherence terms, as detailed in~\cite{rossnagel2015fast}. One kind considers the allowed spontaneous decay channels of the two dipolar transitions characterized by their natural linewidths $\Gamma_{SP}$ and $\Gamma_{DP}$. The other type of decoherence channel will take into account the finite laser linewidths $\Gamma_{\textrm{dop}}$ and $\Gamma_{\textrm{pr}}$. Here we can consider the effects of a finite temperature $T$ by quadratically adding an extra Doppler broadening $\Gamma_T = \abs{\vec{k}_{\textrm{dop}}-\vec{k}_{\textrm{pr}}}\sqrt{\frac{k_B T}{2m}}$ to the laser linewidths.
All operators of the Lindblad operator are proportional to the square root of the linewidths to act as true damping non-unitary terms.

Since the equations are linear in $\rho$, we can rewrite the master equation as
\begin{equation}
    \frac{d\Vec{\rho}}{dt} = L_0\Vec{\rho},
    \label{liouv}
\end{equation}
where $\vec{\rho}$ is the flattened version of the density matrix $\vec{\rho}~=~(\rho_{11},\rho_{12},...,\rho_{18},\rho_{21},...,\rho_{88})$ and $L_0$ is the \textit{Liouvillian} of the system. To obtain the atomic spectra, one can solve the differential equations that arise, named \textit{optical Bloch equations} (OBE), and compute a stationary value varying the detuning of interest. But the computational costs to pay are high, especially if we want to fit a measured spectrum to retrieve experimental parameters. Instead, a stationary solution of Eq. \eqref{liouv} can be found by inverting $L_0$ with the normalization condition Tr$(\rho) = 1$. The spectra, then, can be calculated computing the population of the excited states $\rho_{33}+\rho_{44}$, which is proportional to the fluorescence, while varying one of the frequencies of the fields, keeping the other one constant. All the spectra of Fig. \ref{fig:01_cpttheo} except the dashed one of \ref{fig:01_cpttheo}(c) were calculated through this method. In this last one, an additional laser in the $D$-$P$ transition is turned on and the spectrum calculation was performed with the three-laser formalism detailed in the next section.

\subsection{Three-laser CPT theory: Time-averaged solution}
We now consider adding one more laser to the multi-$\Lambda$ system. In particular, we consider adding a second laser nearly resonant to the $D$-$P$ transition near 866~nm with independent frequency with respect to the other one. Its addition to the interaction Hamiltonian in Eq. \eqref{intham} yields a problem: it is not possible to build a unitary transformation $U$ to go into a rotating system in which the Hamiltonian becomes time independent. 
Choosing the same rotation $U$ used in the two-laser case, one obtains a time-dependent Hamiltonian as
\begin{equation}
    \mathcal{H}' = \mathcal{H}_0 + \frac{\hbar \Omega_{\textrm{rep}}}{2} \left[ H_+ e^{i\bar{\Delta} t} +  H_- e^{-i\bar{\Delta} t} \right],
    \label{hmashmenos}
\end{equation}
where $\bar{\Delta} = \Delta_{\textrm{rep}} - \Delta_{\textrm{pr}}$ is the difference between the detuning of the repumper and the probe, $\mathcal{H}_0$ is the Hamiltonian of the two-laser system, $\Omega_{\textrm{rep}}$ is the Rabi frequency of the repumper, and $H_+$ and $H_-$ are two $8 \times 8$ matrices which satisfy the condition $H_+^{\dagger} = H_-$ and depend only on the repumper polarization. It is straightforward to see that in this frame of reference, making $\Omega_{\textrm{rep}} = 0$ directly gives the two-laser equations. 

Using the same procedure as in the previous section, expressing all the elements of the density matrix as a row vector and correspondingly the elements of this time-dependent Hamiltonian as matrix operators, we obtain a new master equation as
\begin{equation}
    \frac{d\Vec{\rho}}{dt} = \left[L_0+ \frac{\hbar \Omega_{\textrm{rep}}}{2}\left( L_{+} e^{i\bar{\Delta}t} + L_{-} e^{-i\bar{\Delta}t}  \right)\right]\Vec{\rho}.
    \label{lmaslmenos}
\end{equation}
Here, $L_0$ is the Liouvillian for the two-laser system of Eq. \eqref{liouv}, and $L_{\pm}$ are two $8^2 \times 8^2$ matrices that are calculated with the expressions of $H_{\pm}$.

As the Liouvillian is now explicitly time-dependent, we cannot find the stationary solutions by simply inverting this matrix as before. However, since it is periodic with frequency $\bar{\Delta}$, the stationary solution will also have the same periodicity. Because of this, we propose a Floquet-type solution of the form $\vec{\rho}(t) = \sum_{n}\vec{\rho}_n e^{in\bar{\Delta}t}$. 
By introducing it into the master equation, we get a recursive expression of the form
\begin{equation}
    \left[L_0 - in\bar{\Delta}\right]\vec{\rho}_n + \frac{\hbar \Omega_{\textrm{rep}}}{2}L_{+}\vec{\rho}_{n-1} + \frac{\hbar \Omega_{\textrm{rep}}}{2}L_{-}\vec{\rho}_{n+1} = 0.
    \label{Ecrecursiva}
\end{equation}
Defining two lowering and raising operators in a recursive form as
\begin{equation}
    S^{+}_{n-1} = -\left(L_0 -in\bar{\Delta} + \frac{\hbar \Omega_{\textrm{rep}}}{2}L_{-}S_n^{+}\right)^{-1}\frac{\hbar \Omega_{\textrm{rep}}}{2}L_{+},
    \label{Splus}
\end{equation}
\begin{equation}
    S^{-}_{n+1} = -\left(L_0 -in\bar{\Delta} + \frac{\hbar \Omega_{\textrm{rep}}}{2}L_{+}S_n^{-}\right)^{-1}\frac{\hbar \Omega_{\textrm{rep}}}{2}L_{-},
    \label{Sminus}
\end{equation}
we can write $\vec{\rho}_{n+1} = S_n^{+}\vec{\rho}_n$ for $n \geq 0$ and $\vec{\rho}_{n-1}=S_n^{-}\vec{\rho}_n$ for $n \leq 0$. 

Of  all of the $\vec{\rho}_n$ terms we will be interested only in finding the mean value $\vec{\rho}_0 = \bar{\Delta}\int_0^{1/\bar{\Delta}}\vec{\rho}(t)dt $ as this is what we measure in the experiments.
Using these expressions, according to Eq. \eqref{Ecrecursiva} we can get an expression for $\vec{\rho}_0$ of the form
\begin{equation}
    \left(L_0 + \frac{\hbar \Omega_{\textrm{rep}}}{2}L_{-}S_0^{+} + \frac{\hbar \Omega_{\textrm{rep}}}{2}L_{+}S_0^{-} \right)\vec{\rho_0} = 0.
    \label{Liouvillemodificado}
\end{equation}
This shows that the extra laser in the $D$-$P$ transition adds a correction to the Eq. \eqref{liouv} which depends on the $L_{\pm}$ matrices via the recursive relations for $S_0^{\pm}$. These can be found iteratively by setting an upper and lower bound to the sum as $ S^{+}_{n_{\textrm{max}}} = S^{-}_{-n_{\textrm{max}}} = 0$, which is analog to truncating the sum of $\vec{\rho}(t)$. A good choice for the cut of the sum is $n_{\textrm{max}}=\frac{\Omega_{\textrm{rep}}}{\bar{\Delta}}$ since it is the adiabatic elimination parameter of the $\Lambda$-system formed by two sublevels D and one P, hence characterizing the oscillations. For most of the spectrum, $\frac{\Omega_{\textrm{rep}}}{\bar{\Delta}}\ll1$ so setting $n_{\textrm{max}}=1$ will adequately reproduce the expected values except when    $\left| \Delta_{\textrm{rep}} - \Delta_{\textrm{pr}} \right| \sim \Omega_{\textrm{rep}}$, where a bigger value of $n_{\textrm{max}}$ has to be used. We see that choosing $n_{\textrm{max}}=5$ is sufficient to reproduce all observed values within typical experimental error.

\begin{figure}[b]
\includegraphics[scale=1]{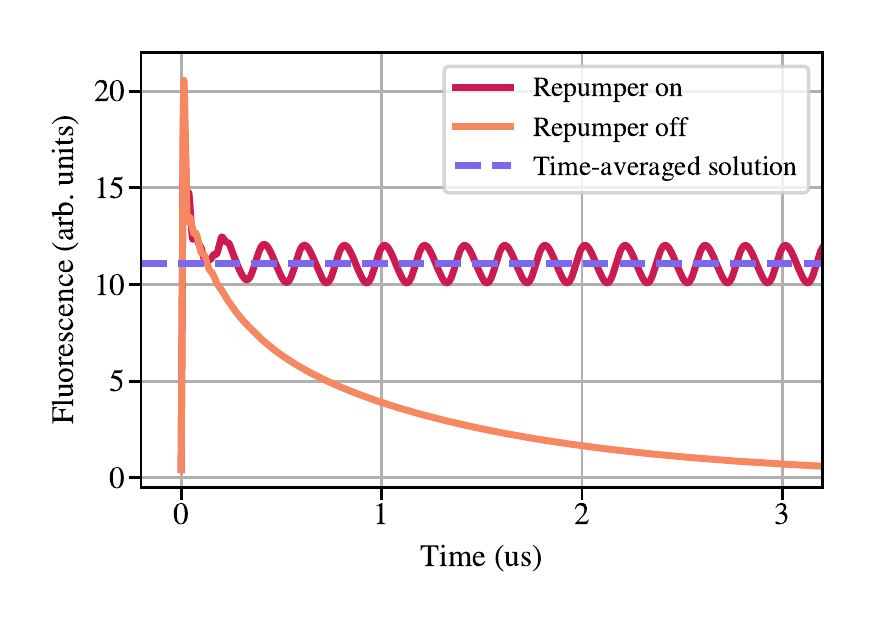}
\caption{\label{fig:sims} Simulated time-dependent fluorescence, proportional to the excited state population, for a single ion with and without an extra repumper for a configuration with an open decay pathway. When there is no repumper (orange solid line) the fluorescence tends to 0 after a transient time due to optical pumping to the $m_j = \pm 3/2$ magnetic sublevels of the $D$ state. By adding an extra laser in the $D$-$P$ transition (repumper) with $\hat{\sigma}_+ + \hat{\sigma}_-$ polarization, the fluorescence oscillates around a non-zero mean value (red solid line). This is well described by the time-averaged solution (purple dashed line) with $n_{\textrm{max}}=3$.}
\end{figure}

With this, we get a time-independent equation for $\vec{\rho}_0$, which can be solved in the same way that the two-laser problem. As a result of this, we have a method to compute a time-averaged solution of the problem, i.e., the stationary value of the fluorescence averaging all the oscillations. This is the case when we perform the experiments with a detector which is not locked to the relative phase of the lasers and do many repetitions, or when the $D$-$P$ lasers' relative coherent time is much smaller than the characteristic experiment times. This last one is the case when two different light sources are used as we do in the experiment.

In Fig. \ref{fig:sims} we can see a comparison between the exact solution of the OBE for two and three lasers, and the time-averaged solution obtained with the previously described formalism. When the polarization of the probe is such that we have optical pumping to some D magnetic sublevels (orange solid line), for example with $\hat{\pi}$ polarization, after a transient, the excited state population, proportional to the fluorescence, tends to 0. But if we turn on an extra laser in the $D$-$P$ transition (red solid line) with polarization such that all sublevels are repumped, like $\hat{\sigma}_+ + \hat{\sigma}_-$, optical pumping is avoided and the mean value of it now is different to zero. We can also see that the excited state population oscillates around the time-averaged solution (purple dashed line), where $n_{\textrm{max}}=3$ was used. This shows that the last one is a suitable approximation to the stationary value of the three-laser problem.

The dashed plot in Fig. \ref{fig:01_cpttheo}(c) was calculated with this method. 
We can see that the pure $\hat{\pi}$ dark resonances can be successfully resolved by adding an additional laser which prevents from optical pumping to the $D$ sublevels. 

In the next sections we will show some applications of the model, as well as two possibilities to do polarimetry of the probe laser studying the dark resonance spectrum.

\section{EXPERIMENTAL RESULTS}

\subsection{Experimental setup}
\begin{figure}[b]
\includegraphics[scale=1]{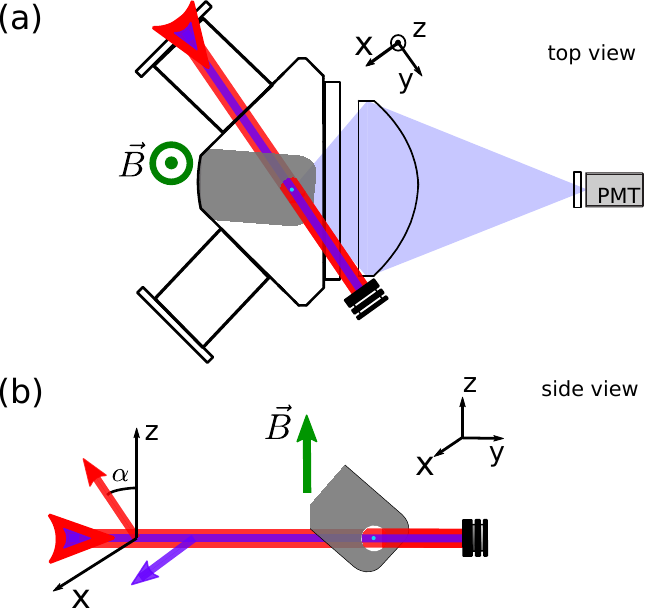}
\caption{\label{fig:setup} Sketch of the experimental setup. (a) A single ion (light blue) is trapped in a ring-shaped Paul trap (gray) placed in a ultra-high vacuum chamber. The fluorescence (pale-violet) is collected by an aspheric objective and focused onto a photomultiplier tube (PMT) with a mounted $(397\pm5)$-nm interference filter. The incident beams are collinear and propagate in the horizontal ($\hat{y}$) direction. The magnetic field $\vec{B}$ is vertically ($\hat{z}$) oriented. (b) The polarization of the beams is linear and lies in the $xz$-plane. The polarization of the 397-nm laser is always orthogonal to $\vec{B}$, and the polarization of both 866-nm lasers forms angles $\alpha_{\textrm{rep}}$ and $\alpha_{\textrm{pr}}$ with respect to the the $\hat{z}$ direction. }
\end{figure}

The experiments were performed with a single $^{40}$Ca$^+$ ion trapped in a ring-shaped Paul trap, as shown in Fig. \ref{fig:setup}. The radio-frequency field was set to $22.1$ MHz and $\sim$ 600 V$_{pp}$. These parameters generate trap frequencies in the range of $0.7-1.5$ MHz.  

Two different lasers near $866$~nm were used to repump and probe the $D$-$P$ transition. These two beams were focused on the ion's position with a beam waist of $\sim 200~\mu$m. Also, the beams were overlapped together with a $397$-nm laser with a beam waist of $\sim 80~\mu$m which addresses the $S$-$P$ transition. The propagation direction of the three lasers form a collinear configuration. The $397$-nm laser and the $866$-nm probe laser are locked to individual Fabry-Perot cavities, with a free spectral range of 1.5~GHz and a finesse of $\sim200$, through the Pound-Drever-Hall method, achieving linewidths below $0.1$~MHz~\cite{luda2019compact}. The $866$-nm repumper laser was a free-running titanium-sapphire laser. In all cases, the beams were linearly polarized, and their propagation direction formed an angle of $130^{\circ}$ with respect to the detection direction. All the lasers were frequency tuned and scanned using several acousto-optic modulators (AOMs) in double-pass configurations. The fluorescence detection was performed with a Hamamatsu H10682 photomultiplier tube (PMT) with a mounted 397-nm optical filter to detect scattered photons only of the $S$-$P$ transition. 
The AOMs and the PMT were controlled and synchronized with an ARTIQ system~\cite{bourdeauducq2016artiq,kasprowicz2020artiq}. 
The magnetic field was generated by a combination of permanent magnets located below the trap and three orthogonal coils. This allowed us to align the field in the vertical direction orthogonal to the propagation direction of the lasers. The magnitude of the field was $\sim$4~G.

\begin{figure}[b]
\includegraphics[scale=1]{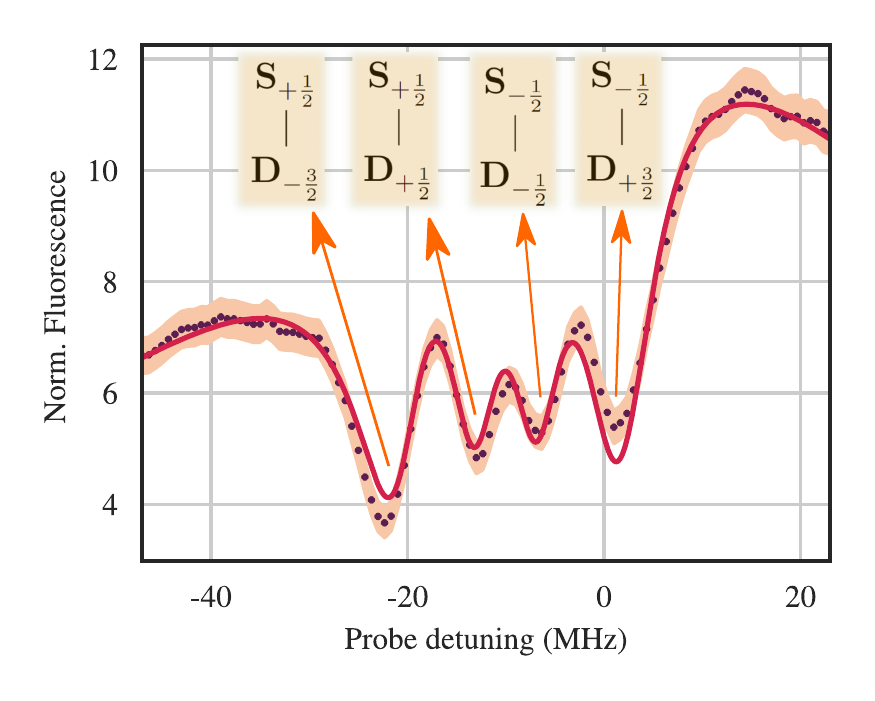}
\caption{\label{fig:cpt2lasers} Dark resonance spectrum with two lasers, where $\hat{\epsilon}_{\textrm{dop}}=\hat{\sigma}_+ + \hat{\sigma}_-$ and $\hat{\epsilon}_{\textrm{pr}}=\hat{\sigma}_+ + \hat{\sigma}_-$. The experimental data (dots) are plotted with their uncertainty (shaded area) along with a fit to the two-laser model (solid line). From the fit, we retrieve $S_{\textrm{dop}}=0.53(1)$, $S_{\textrm{pr}}=9.9(2)$ and $\Delta_{\textrm{dop}}=-9.8(1)~$MHz. With this configuration, four dark resonances appear in the spectrum, each one corresponding to a dark state preparation of a superposition of one $S$ magnetic sublevel and one different $D$ magnetic sublevel.}
\end{figure}

\subsection{Measurement protocol}
The dark resonance spectra were obtained by scanning the probe frequency and measuring the fluorescence on the 397~nm transition in a pulsed scheme. For each measurement point the following sequence was performed. First, the ion is Doppler cooled for $100~\mu$s with a $397$-nm red-detuned and a $866$-nm blue-detuned lasers, both $\hat{\sigma}_+ + \hat{\sigma}_-$ polarized to ensure that the ion is always initialized at the same temperature. Then, the frequencies of the lasers are changed to the desired values. After a waiting time of $20~\mu$s to avoid transient effects, the ion's fluorescence is detected with the PMT for a time interval of $50$~$\mu$s such that no significant heating occurs during the measurement. This is repeated $20.000$ times to reduce statistical uncertainties. Also, once every 50 repetitions the ion is cooled for $2$~ms to avoid excess heating.

\begin{figure*}[htp]
\includegraphics[scale=1]{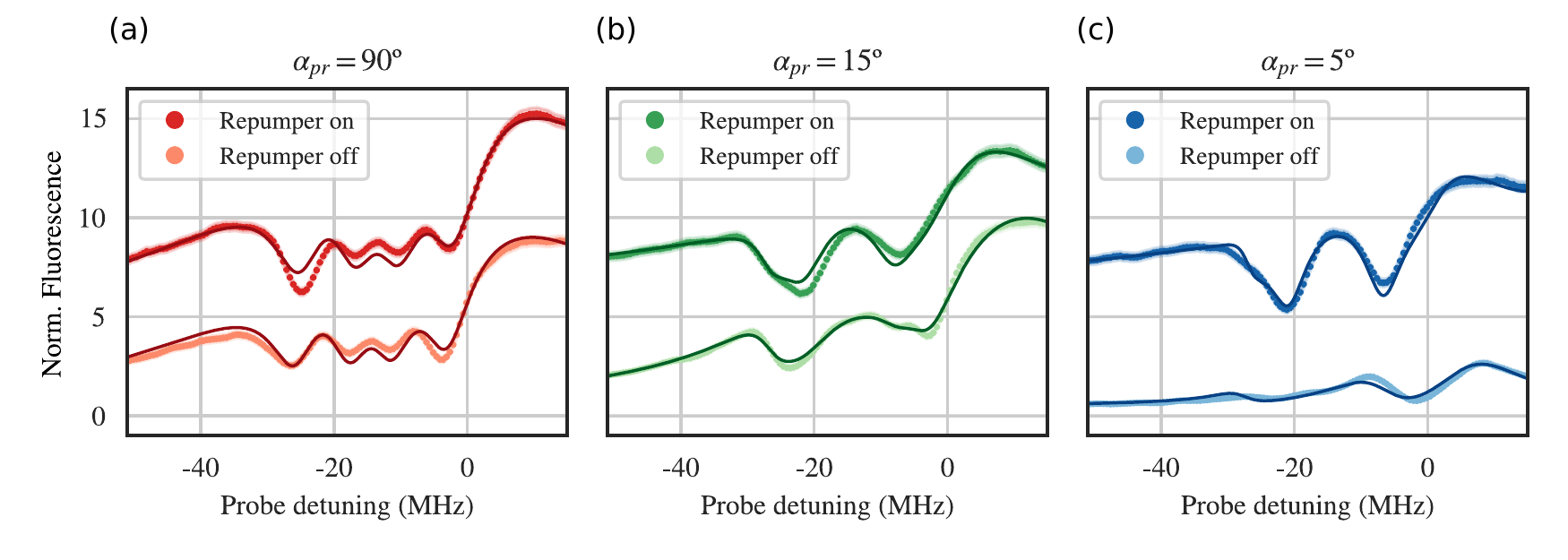}
\caption{\label{fig:sweepingprobe} Dark resonance spectra for three different angles of the probe laser polarization, with a repumper on and off. The measurements (dots) are plotted with their uncertainties (shaded area). In all cases, a fit to the model is shown (solid line) using the two-laser model (repumper off case) and the three-laser model (repumper on case). For all curves, $S_{\textrm{dop}}=0.68(2)$, $S_{\textrm{pr}}=7.6(2)$ and $\Delta_{\textrm{dop}}=-14.7(1)$~MHz. (a) For $\alpha_{\textrm{pr}}=90^{\circ}$, the repumper effect is a general increase of the overall fluorescence. (b) For $\alpha_{\textrm{pr}}=15^{\circ}$, the addition of the repumper  makes the two outer dark resonances merge with two emerging $\hat{\pi}$ dark resonances, which cannot be resolved individually. (c) For $\alpha_{\textrm{pr}}=5^{\circ}$, i.e., close to $0^{\circ}$, without the repumper the spectrum is poorly resolved and little information can be retrieved from it. But with the repumper on, the two $\hat{\pi}$ dark resonances appear. From the fits, for the case where the repumper is on we extract $S_{\textrm{rep}}=\{2.51(11), 2.83(5), 2.72(8)\}$ and $\Delta_{\textrm{rep}}=\{24.8(3), 25.7(2), 26.0(2)\}$~MHz for plots (a) to (c) respectively.}
\end{figure*}

\begin{figure}[h]
\includegraphics[scale=1]{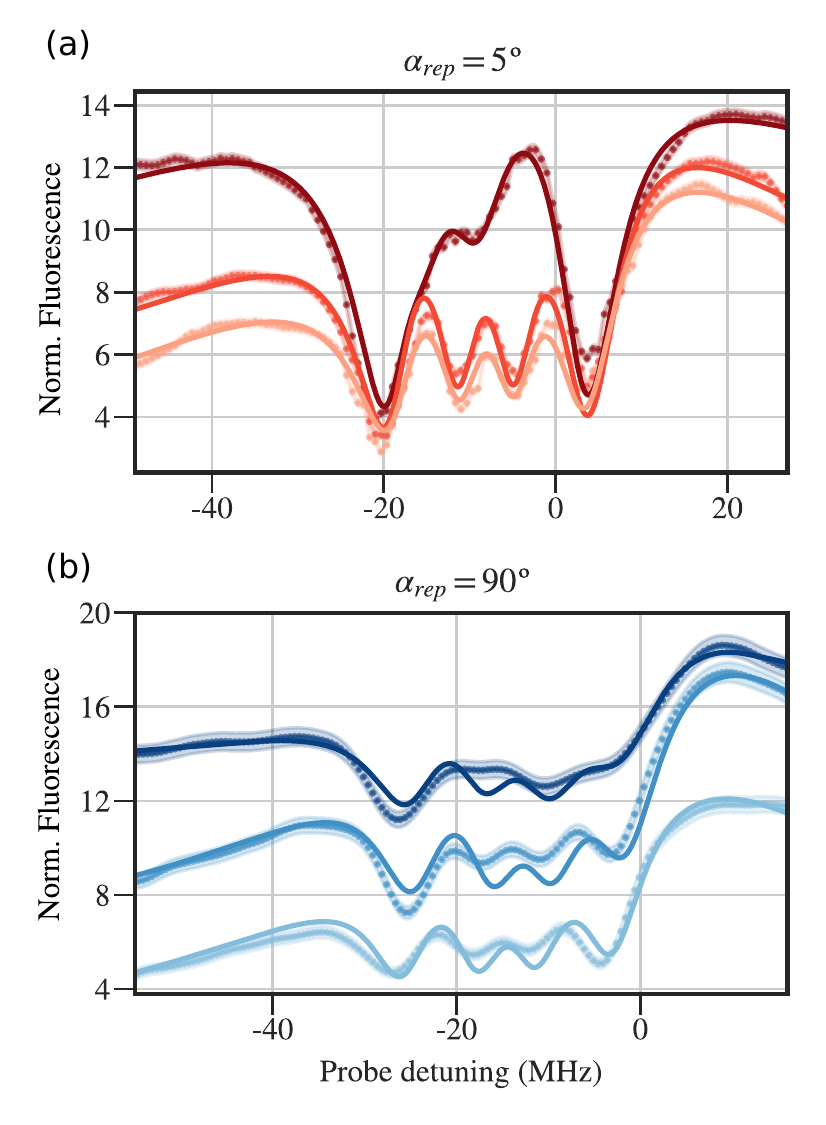}
\caption{\label{fig:drkicking} Dark resonance spectra showing resonance ``kicking" for different laser polarizations and powers. The spectra are analog to the one of Fig. \ref{fig:cpt2lasers}, but with an additional (repumper) laser in the $D$-$P$ transition. The measurements (dots) are plotted with their uncertainties (shaded areas), and along fits to the model (solid lines) presented in the text. In all cases, $\Delta_{\textrm{dop}} = -14.6(2)~$MHz, $S_{\textrm{dop}} = 0.65(2)$, and $S_{\textrm{pr}} = 7.43(16)$. (a) When the repumper is $\hat{\pi}$ polarized, the two middle dark resonances, i.e., the resonances that involve $m_j = \pm 1/2$ $D$ sublevels are kicked, which means that the dark state formed is decohered by the extra laser. Here, the saturation parameters of the repumper extracted from the fits to the model, from the light to the dark curve, are $S_{\textrm{rep}}=\{0, 1.61(5), 7.23(21)\}$. The detuning of the repumper obtained is $\Delta_{\textrm{rep}}=52.3(31)~$MHz. (b) When the repumper is $\hat{\sigma}_+ + \hat{\sigma}_-$ polarized, all four dark resonances are kicked since it repumps all four $D$ magnetic sublevels and hence the four dark states are affected. In this case, the saturation parameters of the repumper obtained from the fits, from the light to the dark curve, are $S_{\textrm{rep}}=\{0, 2.68(11), 9.31(30)\}$, and its detuning is $\Delta_{\textrm{rep}}=30.6(5)~$MHz.}
\end{figure}

\subsection{Typical dark resonance spectrum with two lasers}

In Fig. \ref{fig:cpt2lasers} we show a dark resonance spectrum with the fitted model obtained with two lasers. The polarization of the Doppler laser was set to $\hat{\sigma}_+ + \hat{\sigma}_-$ since it is the polarization that separates the dark resonances the most, and this will be kept through all the work. In this case, the polarization of the probe laser was also set to $\hat{\sigma}_+ + \hat{\sigma}_-$. This is the case exemplified in Fig. \ref{fig:01_cpttheo}(a). As expected, we see the four dark resonances. The two middle ones involve $m_j=\pm1/2$ $D$ magnetic sublevels, and the two outer ones involve $m_j=\pm3/2$ $D$ magnetic sublevels.

By fitting the spectrum for the two-laser configuration, we obtain several crucial experimental parameters. One is the magnetic field, since the separation of the dark resonances is linear with $B$. This yields $B=3.7(1)$ G. Another parameter retrieved from the fit is the temperature of the ion, which in this case is $4.7(4)$ mK. We can also get the saturation parameters $S=I/I_{\textrm{sat}}$ of the lasers or, equivalently, the Rabi frequencies of the transitions. Additionally, we include a scale and an offset factor to account for the conversion between atomic population and photon counts considering also background light.


To measure the dark resonances, we observed that it is important to keep the $S$-$P$ transition low saturated, with $S_{\textrm{dop}} \leq 0.8$ ($\sim 15~\mu$W). On the other hand, we keep the $D$-$P$ transition very saturated, with $S_{\textrm{pr}} \approx 5$ ($\sim 40~\mu$W). This last condition will be reflected in power broadening of the full spectrum but is beneficial to gain contrast in the resonances.  While keeping the parameters in these orders, we proceed to turn on a third laser and analyze some aspects of the three-laser model.

\subsection{Dark resonance spectra with three lasers}
We now present, in Fig. \ref{fig:sweepingprobe}, the measured spectra and fits to our model for three different probe polarizations. Each one is shown with and without a weak repumper laser with polarization $\hat{\epsilon}_{\textrm{rep}}=\hat{\sigma}_+ + \hat{\sigma}_-$, to show its effect on optical pumping.
For the first two cases we see clear resonances both with and without the extra repumper, with $S_{\textrm{rep}} \sim 3$ and $\Delta_{\textrm{rep}} \sim 20$~MHz. Here, the main effect of the third laser is to raise all the spectra by some amount, which is well predicted by the model. 
However, for the last case, when the polarization of the probe is almost purely $\hat{\pi}$, we see that the spectrum is only well defined when the repumper is on. In this case, the extra laser avoids optical pumping into the $m_j=\pm3/2$ $D$ states and allows us to recover the dark resonance spectrum for the two $\hat{\pi}$ transitions, as predicted. 
In effect, this shows that an additional laser in the $D$-$P$ transition results in avoiding optical pumping to $D$ sublevels that are not addressed by the scanning probe laser for certain polarizations. 
Similarly, if the probe had pure $\hat{\sigma}_+$ or $\hat{\sigma}_-$ polarization, a weak repumper laser would allow the observation of other single pairs of dark resonances.

\subsection{Dark resonance kicking}

Contrary to the case of a weak repumper where dark resonances are avoided, we now consider the case when its power is comparable to the one of the probe laser. Now the dark resonance structure generated by the probe and the Doppler lasers may be altered because the repumper can break the coherences of the dark states. 
In this case, the probe and the repumper lasers compete by generating and breaking the coherences through photon scattering.
An illustration of this effect is seen in Fig. \ref{fig:drkicking}.
There, the probe is set to have $\hat{\sigma}_+ + \hat{\sigma}_-$ polarization generating four dark resonances in the absence of a repumper laser. Each of these resonances corresponds to a superposition of only one of the different $D$ sublevels with one of the levels in the $S$ manifold, as also shown in Figs. \ref{fig:01_cpttheo}(a) and \ref{fig:cpt2lasers}. 
Now, as the repumper is turned on, and its power becomes comparable to the one of the probe, we see that some of the resonances lose contrast. They get ``kicked" by the repumper. In Fig. \ref{fig:drkicking}(a) the repumper has $\hat{\pi}$ polarization so it kicks the dark resonances that involve $m_j=\pm1/2$ $D$ magnetic sublevels, which are the middle ones. Conversely, when the repumper laser has $\hat{\sigma}_+ + \hat{\sigma}_-$ polarization, as shown in Fig. \ref{fig:drkicking}(b), all the dark resonances are equally smeared out.

We see that the grade to which the repumper kicks the different resonances reveals its polarization. In particular, we will now show that it can be used to determine the angle between the linear polarization of this laser and the magnetic field. In the more general case, by varying the magnetic field, one can measure all three components of a vectorial polarization.
Therefore, if we can quantify how much the repumper kicks the sides resonances versus the middle resonances we can retrieve the proportion of $\hat{\pi}$ and the proportion of $\hat{\sigma}_+ + \hat{\sigma}_-$ of polarization it carries. Consequently, the quantification of the dark resonances kicking could be used as an estimator of the angle between the repumper laser and $\vec{B}$.

\section{POLARIMETRY WITH CPT SPECTRA}
We now consider the two limits described above to be applied to polarimetry methods. We assume that the two lasers that address the $D$-$P$ transition near $866~$nm have linear polarization and their propagations are perpendicular to $\vec{B}$. We want to build an estimator of the polarization of one of the lasers knowing the other one by using dark resonance spectra. We aim at retrieving the angle $\alpha$ between the polarization of the laser of interest and $\vec{B}$. 

Here, we propose two methods for one of those lasers. The first method consists in analyzing the full dark resonance spectrum of the probe laser using a weak blue-detuned repumper to determine the polarization of the probe by the depths of the observed resonances, as in Fig. \ref{fig:sweepingprobe}. 
In the second method we quantify the polarization direction of the repumper laser by analyzing the degree to which the different dark resonances are ``kicked" for a fixed configuration of probe and Doppler laser, as in Fig. \ref{fig:drkicking}.

\subsection{Scanning probe polarimetry}
Here, the aim is to determine the polarization of the probe laser. We turn on a weak blue-detuned repumper laser to be able to observe the full dark resonance spectra, as we showed before. As seen in Fig. \ref{fig:sweepingprobe}, the number and the depths of the dark resonances strongly depend on $\alpha_{\textrm{pr}}$. The two dark resonances that emerge for $\alpha_{\textrm{pr}}=5^{\circ}$ are in the middle of the four $\hat{\sigma}_+ + \hat{\sigma}_-$ dark resonances that appear for $\alpha_{\textrm{pr}}=90^{\circ}$. This shows that we can observe dark resonance spectra for every probe polarization without optical pumping.

To retrieve the polarization from these spectra, two limits are considered. If the magnetic field is high enough such that the dark resonances can be distinguished separately, one could obtain information of the polarization of the beam characterizing the depths of each dark resonance. In the other case, if the magnetic field does not separate enough the resonances, this method relies on the fit to the model of the measured spectrum.

\subsection{Dark resonances kicking polarimetry}
Now we consider the use of the resonance kicking effect on a fixed reference dark resonance spectrum. This has two advantages over the previous method: first, we have a reference spectrum setting $S_{\textrm{rep}}=0$ from which we can get all experimental parameters that do not involve the laser of interest; and second, since we only need to be able to see four dark resonances with the scanning laser, this method is more robust regarding the parameters space in which this works. 

\begin{figure}[h]
\includegraphics[scale=1]{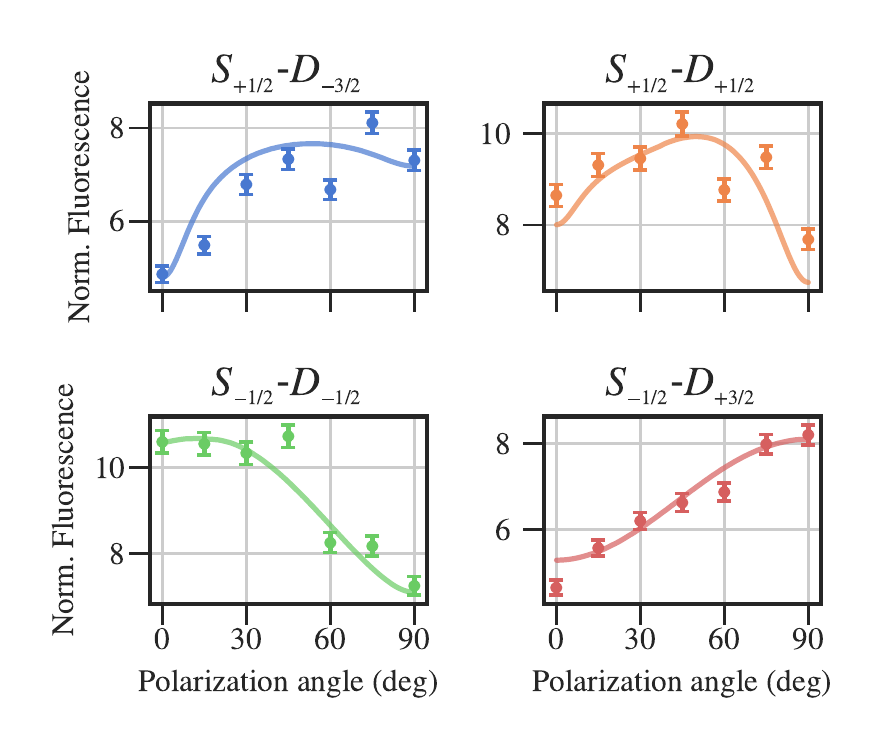}
\caption{\label{fig:polarimetersensibility} Dark resonances ``kicking" and their dependence on the beam's polarization. Each curve shows the fluorescence measured at the frequency corresponding to each of the dark resonances shown in Fig. \ref{fig:cpt2lasers} as a function of the polarization angle of the probe beam with respect to the magnetic field $\vec{B}$. The repumper was set to $S_{\textrm{rep}}=6.22(14)$ and $\Delta_{\textrm{rep}}=32.9(3)~$MHz, and the Doppler parameters used were $S_{\textrm{dop}}=0.61(2)$ and $\Delta_{\textrm{dop}}=-9.6(1)$~MHz. Also, for this measurements, $S_{\textrm{pr}}=4.83(5)$. The dots are retrieved from the measured curves. The solid lines represent theoretical curves simulated with fitted parameters extracted from two reference spectra.}
\end{figure}

To quantify the polarization angle we measure the depths of each of the four dark resonances for a fixed strong repumper, while varying its angle with respect to the magnetic field $\vec{B}$. In Fig. \ref{fig:polarimetersensibility} we see the results along with theoretical calculations. These were obtained by simulating spectra with parameters retrieved from two fits: a first fit of the spectrum with two lasers to obtain all parameters that do not involve the repumper laser, and a second fit of one spectrum with three lasers fixing $\hat{\epsilon}_{\textrm{rep}}$ to get $S_{\textrm{rep}}$ and $\Delta_{\textrm{rep}}$. We can see that the experimental data are in good agreement with the simulations. Despite the non-monotonic behavior of two of the dark resonances, the mapping between $\alpha_{\textrm{rep}}$ and the depths of the dark resonances can be successfully made, proving to be a feasible method to do polarimetry of a linearly polarized repumper laser.

This method can also be extended to characterize circularly or elliptically polarized beams. The difference would rely on the fact that the pair of dark resonances that are decohered are different. For example, if the repump laser is $\hat{\sigma}_+$ polarized, the dark resonances that would be kicked are one of the middle and one of the side instead of the two middle ones. And, by decomposing the polarization in $\hat{\pi}$, $\hat{\sigma}_+$, and $\hat{\sigma}_-$, one can identify the pair of dark resonances that should be ``kicked" and by quantifying this effect one could retrieve information about the beam. 

This method can be applied, for example, to characterize the polarization profile of a highly focused vector beam~\cite{quinteiro2017twisted,wang2020vectorial}. This kind of beams shows a counter-intuitive behavior when focused.
For example, radially polarized beams have, when focused close to the diffraction limit, strong longitudinal fields exceeding their transverse component. 
This only becomes significant when the beams are focused to waists below $5~\mu$m, requiring nano-sized sensors to measure the spatial distribution of the polarization of these beams. 
A single trapped ion, with typical wave packets in the $10$-$100$~nm range, is then particularly well suited for this task and would allow measuring the full polarization profile of these focused beams.

\section{$3^2D_{3/2}-3^2D_{3/2}$ DARK RESONANCES AND APPLICATION TO THERMOMETRY}
\begin{figure}[h]
\includegraphics[scale=1]{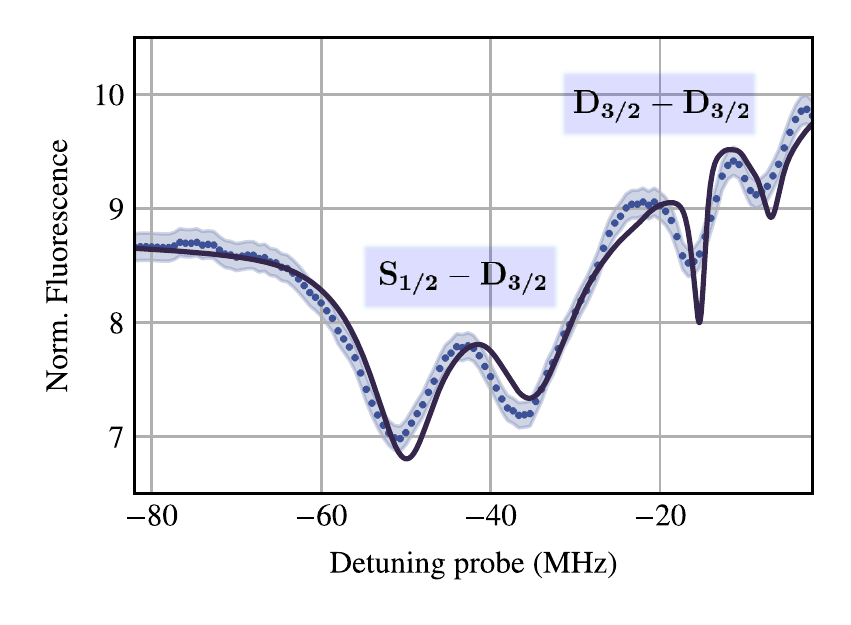}
\caption{\label{fig:drsublevels} Dark resonance spectrum with three lasers. 
On the right, the two dark resonances generated between the $3^2D_{3/2}-3^2D_{3/2}$ levels are visible. These are sharper than the two resonances on the left which correspond to the well-known $4^2S_{1/2}-3^2D_{3/2}$ dark states which we analyzed in detail in this paper.
For this measurement the laser polarizations were
$\hat{\epsilon}_{\textrm{dop}}=\hat{\sigma}_+ + \hat{\sigma}_-$, $\hat{\epsilon}_{\textrm{pr}}=\hat{\pi}$ and $\hat{\epsilon}_{\textrm{rep}}=\hat{\sigma}_+ + \hat{\sigma}_-$. In this case, $S_{\textrm{dop}}=1.17(1)$, $S_{\textrm{pr}}=7.61(23)$, $S_{\textrm{rep}}=14.0(3)$, and the detunings are $\Delta_{\textrm{dop}}=-42.3(5)$~MHz and $\Delta_{\textrm{rep}}=-11.4(2)$~MHz.
}
\end{figure}

The previous results were obtained by blue-detuning the repumper out of the probe scanning range to avoid possible mutual interference effects between the two lasers that address the $D$-$P$ transition. When the detuning of these two lasers match, additional dark resonances appear generated by coherent superpositions of magnetic sublevels in the $3^2D_{3/2}$ manifold. Here, again, the polarizations of both lasers determine the strength of the observed resonances. 
In Fig. \ref{fig:drsublevels} we show an example where two $3^2D_{3/2}-3^2D_{3/2}$ dark resonances as well as two $4^2S_{1/2}-3^2D_{3/2}$ are seen. 
Here we set $\hat{\epsilon}_{\textrm{pr}}=\hat{\pi}$ and $\hat{\epsilon}_{\textrm{rep}}=\hat{\sigma}_+ + \hat{\sigma}_-$. We note that the theory adequately describes the $4^2S_{1/2}-3^2D_{3/2}$ transitions, while it predicts sharper $3^2D_{3/2}-3^2D_{3/2}$. We attribute this to frequency drifts of the repumper laser which was not locked to a reference cavity, as the other two were. Also, we note that, if both repumper and probe laser were either derived from the same laser, or frequency locked, the contrast of these resonances would be even sharper, and independent of laser drifts or linewidths.

An interesting feature of the observed resonances is that, as the propagation of both lasers is parallel and the wavelengths are very close to each other, the Doppler broadening of the $3^2D_{3/2}-3^2D_{3/2}$ dark resonances is negligible, so these are not at all sensitive to the ion's temperature.

Nevertheless, by changing the angle between both $866$~nm lasers, those dark resonances can become temperature sensitive. 
Varying the propagation direction of the repumper in the plane orthogonal to $\vec{B}$ would then tune the sensitivity of these dark resonances to the ion's temperature, i.e., the range of measurable temperatures with the method. Such a method would have the possibility of tuning sensitivity with angle as  in~\cite{rossnagel2015fast} and the independence of the laser linewidths when both probe and repumper are derived from the same source and tuned via two independent AOMs as in~\cite{tugaye2019absolute}.

\section{CONCLUSIONS}
In this work we studied the dark resonance spectra in a multilevel-$\Lambda$ system with three lasers, particularly the $^{40}$Ca$^+$ eight-level system.
We showed both theoretically and experimentally that including an extra laser to the standard two-laser configuration allows measuring CPT spectra that otherwise would be impossible to observe, avoiding optical pumping out of the fluorescence cycle. We also proved that this additional laser can break the coherences of dark states which modify the spectra. Since this effect strongly depends on the polarization of this laser, a method for performing 3D polarimetry of highly focused vector beams can be implemented. Finally, we showed the emergence of additional $3^2D_{3/2}-3^2D_{3/2}$ dark resonances that represent dark states composed of Zeeman sublevels of the $D$ state, and discussed how these can be used to perform ion thermometry with tunable sensitivity and independently of the laser linewidths.

\section*{ACKNOWLEDGEMENTS}
We thank F. Schmidt-Kaler for his support and generosity as well as J. P. Paz, M. A. Latoronda, and A. J. Roncaglia for their unconditional help and backup in the setting up of the laboratory. Finally, we thank C. Cormick for carefully reading and making suggestions on the manuscript and M. Ingouville for English proof reading.
This work was supported by Agencia I+D+i Grants No. PICT 2018 - 3350 and No. PICT 2019 - 4349, Secretaría de Ciencia y Técnica, Universidad de Buenos Aires (20020170100616BA) Grant No. UBACyT 2018, and CONICET (Argentina). 

\section*{APPENDIX: CONSTANT MATRICES EXPRESSIONS}
We detail some expressions for the calculation of the Liouvillian in the three-laser case.

The Hamiltonian of Eq. \eqref{hmashmenos} considers two $8 \times 8$ matrices $H_\pm$ that depend only on the polarization of the repumper laser. These fulfill the condition $H_- = H_+^{\dagger}$ with
\begin{equation}H_+ = 
\begin{pmatrix}
0 & 0 & 0 & 0 & \cdots & 0\\
\vdots & \vdots & \vdots & \vdots & \ddots & \vdots\\
0 & 0 & \frac{1}{2}\sin \theta_r e^{i\phi_r} & 0 & \cdots & 0\\
0 & 0 & -\frac{1}{\sqrt{3}}\cos \theta_r & -\frac{1}{2\sqrt{3}}\sin \theta_r e^{i\phi_r} & \cdots & 0\\
0 & 0 & \frac{1}{2\sqrt{3}}\sin \theta_r e^{-i\phi_r} & -\frac{1}{\sqrt{3}}\cos \theta_r & \cdots & 0\\
0 & 0 & 0 & \frac{1}{2}\sin \theta_r e^{-i\phi_r} & \cdots & 0\\
\end{pmatrix}
\end{equation}
where $\theta_r$ and $\phi_r$ are the polar and azimuthal angles of the linear polarization of the repumper laser with respect to the $\vec{B}$ direction.
Using this, the expressions of the $8^2 \times 8^2$ matrices $L_{\pm}$ of Eq. \eqref{lmaslmenos} written in index notation are
\begin{equation}
    L_{\pm_{8\cdot(r-1)+s, 8\cdot(k-1)+j}}=-\frac{i}{\hbar}(H_{\pm_{r,k}}\delta_{_{j,s}}-H_{\pm_{j,s}}\delta_{_{r,k}}),
\end{equation}
where $\delta_{_{i,j}}$ represents the Kroenecker delta function which is $1$ if $i=j$ and $0$ otherwise.

\bibliography{main_paper}
\end{document}